\documentclass[twocolumn]{aastex62}
\received{}
\revised{}
\accepted{}
\submitjournal{AAS journals}

\shorttitle{Subaru High-$z$ Exploration of Low-Luminosity Quasars (SHELLQs) XIX}
\shortauthors{Matsuoka et al.}

\begin{document}

\title{Quasar Luminosity Function at $z = 7$}

\correspondingauthor{Yoshiki Matsuoka}
\email{yk.matsuoka@cosmos.ehime-u.ac.jp}

\author[0000-0001-5063-0340]{Yoshiki Matsuoka}
\affil{Research Center for Space and Cosmic Evolution, Ehime University, Matsuyama, Ehime 790-8577, Japan.}

\author[0000-0003-2984-6803]{Masafusa Onoue}
\affil{Kavli Institute for Astronomy and Astrophysics, Peking University, Beijing 100871, P.R.China}
\affil{Kavli Institute for the Physics and Mathematics of the Universe, WPI, The University of Tokyo, Kashiwa, Chiba 277-8583, Japan.}
\altaffiliation{Kavli Astrophysics Fellow}

\author[0000-0002-4923-3281]{Kazushi Iwasawa}
\affil{ICREA and Institut de Ci{\`e}ncies del Cosmos, Universitat de Barcelona, IEEC-UB, Mart{\'i} i Franqu{\`e}s, 1, 08028 Barcelona, Spain.}

\author[0000-0002-0106-7755]{Michael A. Strauss}
\affil{Department of Astrophysical Sciences, Princeton University, Peyton Hall, Princeton, NJ 08544, USA.}

\author[0000-0003-3954-4219]{Nobunari Kashikawa}
\affil{Department of Astronomy, School of Science, The University of Tokyo, Tokyo 113-0033, Japan.}

\author[0000-0001-9452-0813]{Takuma Izumi}
\affil{National Astronomical Observatory of Japan, Mitaka, Tokyo 181-8588, Japan.}

\author[0000-0002-7402-5441]{Tohru Nagao}
\affil{Research Center for Space and Cosmic Evolution, Ehime University, Matsuyama, Ehime 790-8577, Japan.}

\author[0000-0001-6186-8792]{Masatoshi Imanishi}
\affil{National Astronomical Observatory of Japan, Mitaka, Tokyo 181-8588, Japan.}
\affil{Department of Astronomical Science, Graduate University for Advanced Studies (SOKENDAI), Mitaka, Tokyo 181-8588, Japan.}

\author[0000-0002-2651-1701]{Masayuki Akiyama}
\affil{Astronomical Institute, Tohoku University, Aoba, Sendai, 980-8578, Japan.}

\author[0000-0002-0000-6977]{John D. Silverman}
\affil{Kavli Institute for the Physics and Mathematics of the Universe, WPI, The University of Tokyo, Kashiwa, Chiba 277-8583, Japan.}

\author{Naoko Asami}
\affil{Seisa University, Hakone-machi, Kanagawa, 250-0631, Japan.}

\author[0000-0003-2759-5764]{James Bosch}
\affil{Department of Astrophysical Sciences, Princeton University, Peyton Hall, Princeton, NJ 08544, USA.}

\author[0000-0002-6174-8165]{Hisanori Furusawa}
\affil{National Astronomical Observatory of Japan, Mitaka, Tokyo 181-8588, Japan.}

\author{Tomotsugu Goto}
\affil{Institute of Astronomy and Department of Physics, National Tsing Hua University, Hsinchu 30013, Taiwan.}

\author{James E. Gunn}
\affil{Department of Astrophysical Sciences, Princeton University, Peyton Hall, Princeton, NJ 08544, USA.}

\author[0000-0002-6047-430X]{Yuichi Harikane}
\affil{Institute for Cosmic Ray Research, The University of Tokyo, Kashiwa, Chiba 277-8582, Japan.}

\author[0000-0002-1207-1979]{Hiroyuki Ikeda}
\affil{National Institute of Technology, Wakayama College, Gobo, Wakayama 644-0023, Japan.}

\author[0000-0001-9840-4959]{Kohei Inayoshi}
\affil{Kavli Institute for Astronomy and Astrophysics, Peking University, Beijing 100871, P.R.China}

\author[0000-0002-2134-2902]{Rikako Ishimoto}
\affil{Department of Astronomy, School of Science, The University of Tokyo, Tokyo 113-0033, Japan.}

\author[0000-0002-3866-9645]{Toshihiro Kawaguchi}
\affil{Department of Economics, Management and Information Science, Onomichi City University, Onomichi, Hiroshima 722-8506, Japan.}

\author[0000-0003-3214-9128]{Satoshi Kikuta}
\affil{Center for Computational Sciences, University of Tsukuba, Tsukuba, Ibaraki 305-8577, Japan.}

\author[0000-0002-4052-2394]{Kotaro Kohno}
\affil{Institute of Astronomy, The University of Tokyo, Mitaka, Tokyo 181-0015, Japan.}
\affil{Research Center for the Early Universe, University of Tokyo, Tokyo 113-0033, Japan.}

\author[0000-0002-3852-6329]{Yutaka Komiyama}
\affil{National Astronomical Observatory of Japan, Mitaka, Tokyo 181-8588, Japan.}
\affil{Department of Astronomical Science, Graduate University for Advanced Studies (SOKENDAI), Mitaka, Tokyo 181-8588, Japan.}

\author[0000-0003-1700-5740]{Chien-Hsiu Lee}
\affil{W. M. Keck Observatory, Kamuela, HI 96743, USA}

\author[0000-0003-1666-0962]{Robert H. Lupton}
\affil{Department of Astrophysical Sciences, Princeton University, Peyton Hall, Princeton, NJ 08544, USA.}

\author[0000-0002-2933-048X]{Takeo Minezaki}
\affil{Institute of Astronomy, The University of Tokyo, Mitaka, Tokyo 181-0015, Japan.}

\author[0000-0002-1962-904X]{Satoshi Miyazaki}
\affil{National Astronomical Observatory of Japan, Mitaka, Tokyo 181-8588, Japan.}
\affil{Department of Astronomical Science, Graduate University for Advanced Studies (SOKENDAI), Mitaka, Tokyo 181-8588, Japan.}

\author[0000-0001-5769-9471]{Hitoshi Murayama}
\affil{Kavli Institute for the Physics and Mathematics of the Universe, WPI, The University of Tokyo, Kashiwa, Chiba 277-8583, Japan.}
\affil{Department of Physics, University of California, Berkeley, CA 94720, USA.}
\affil{Ernest Orlando Lawrence Berkeley National Laboratory, Berkeley, CA 94720, USA.}

\author[0000-0002-6109-2397]{Atsushi J. Nishizawa}
\affil{Institute for Advanced Research, Nagoya University, Furo-cho, Chikusa-ku, Nagoya 464-8602, Japan.}

\author[0000-0003-3484-399X]{Masamune Oguri}
\affil{Center for Frontier Science, Chiba University, 1-33 Yayoi-cho, Inage-ku, Chiba 263-8522, Japan}
\affil{Department of Physics, Graduate School of Science, Chiba University, 1-33 Yayoi-Cho, Inage-Ku, Chiba 263-8522, Japan}

\author[0000-0001-9011-7605]{Yoshiaki Ono}
\affil{Institute for Cosmic Ray Research, The University of Tokyo, Kashiwa, Chiba 277-8582, Japan}

\author{Taira Oogi}
\affil{Research Center for Space and Cosmic Evolution, Ehime University, Matsuyama, Ehime 790-8577, Japan.}

\author[0000-0002-1049-6658]{Masami Ouchi}
\affil{National Astronomical Observatory of Japan, Mitaka, Tokyo 181-8588, Japan.}
\affil{Institute for Cosmic Ray Research, The University of Tokyo, Kashiwa, Chiba 277-8582, Japan}
\affil{Kavli Institute for the Physics and Mathematics of the Universe, WPI, The University of Tokyo, Kashiwa, Chiba 277-8583, Japan.}

\author[0000-0003-0511-0228]{Paul A. Price}
\affil{Department of Astrophysical Sciences, Princeton University, Peyton Hall, Princeton, NJ 08544, USA.}

\author[0000-0001-6401-723X]{Hiroaki Sameshima}
\affil{Institute of Astronomy, The University of Tokyo, Mitaka, Tokyo 181-0015, Japan.}

\author{Naoshi Sugiyama}
\affil{Kavli Institute for the Physics and Mathematics of the Universe, WPI, The University of Tokyo, Kashiwa, Chiba 277-8583, Japan.}
\affil{Graduate School of Science, Nagoya University, Furo-cho, Chikusa-ku, Nagoya 464-8602, Japan.}

\author{Philip J. Tait}
\affil{Subaru Telescope, Hilo, HI 96720, USA.}

\author[0000-0002-5578-6472]{Masahiro Takada}
\affil{Kavli Institute for the Physics and Mathematics of the Universe, WPI, The University of Tokyo, Kashiwa, Chiba 277-8583, Japan.}

\author[0000-0003-3769-6630]{Ayumi Takahashi}
\affil{Graduate School of Science and Engineering, Ehime University, Matsuyama, Ehime 790-8577, Japan.}

\author[0000-0002-6592-4250]{Tadafumi Takata}
\affil{National Astronomical Observatory of Japan, Mitaka, Tokyo 181-8588, Japan.}
\affil{Department of Astronomical Science, Graduate University for Advanced Studies (SOKENDAI), Mitaka, Tokyo 181-8588, Japan.}

\author[0000-0002-5011-5178]{Masayuki Tanaka}
\affil{National Astronomical Observatory of Japan, Mitaka, Tokyo 181-8588, Japan.}
\affil{Department of Astronomical Science, Graduate University for Advanced Studies (SOKENDAI), Mitaka, Tokyo 181-8588, Japan.}

\author[0000-0002-3531-7863]{Yoshiki Toba}
\affiliation{National Astronomical Observatory of Japan, Mitaka, Tokyo 181-8588, Japan.}
\affil{Institute of Astronomy and Astrophysics, Academia Sinica, Taipei, 10617, Taiwan.}
\affiliation{Research Center for Space and Cosmic Evolution, Ehime University, Matsuyama, Ehime 790-8577, Japan.}

\author[0000-0001-6491-1901]{Shiang-Yu Wang}
\affil{Institute of Astronomy and Astrophysics, Academia Sinica, Taipei, 10617, Taiwan.}

\author[0000-0002-4999-9965]{Takuji Yamashita}
\affil{National Astronomical Observatory of Japan, Mitaka, Tokyo 181-8588, Japan.}



\begin{abstract}
We present the quasar luminosity function (LF) at $z = 7$, measured with 35 spectroscopically confirmed quasars at $6.55 < z <  7.15$.
The sample of 22 quasars from the Subaru High-$z$ Exploration of Low-Luminosity Quasars (SHELLQs) project,
combined with 13 brighter quasars in the literature, covers an unprecedentedly wide range of rest-frame ultraviolet magnitudes over $-28 < M_{1450} < -23$.
We found that the binned LF flattens significantly toward the faint end populated by the SHELLQs quasars. 
A maximum likelihood fit to a double power-law model has a break magnitude $M^*_{1450} = -25.60^{+0.40}_{-0.30}$, 
a characteristic density $\Phi^* = 1.35^{+0.47}_{-0.30}$ Gpc$^{-3}$ mag$^{-1}$, and a bright-end slope $\beta = -3.34^{+0.49}_{-0.57}$, when the faint-end slope is fixed to $\alpha = -1.2$ as observed at $z \le 6$.
The overall LF shape remains remarkably similar from $z = 4$ to $7$, while the amplitude decreases substantially toward higher redshifts, with a clear indication of an accelerating decline at $z \ge 6$.
The estimated ionizing photon density, $10^{48.2 \pm 0.1}$ s$^{-1}$ Mpc$^{-3}$, is less than 1 \% of the critical rate to keep the intergalactic medium ionized at $z = 7$, and thus indicates that quasars are not a major contributor to cosmic reionization.

\end{abstract}


\section{Introduction} \label{sec:intro}

We are witnessing rapid development in the quest for the most distant quasars, driven primarily by deep wide-field surveys at red-optical and near-infrared (IR) wavelengths.
The first discoveries of high-$z$ ($z \ge 6$) quasars were achieved by the Sloan Digital Sky Survey \citep[SDSS; e.g.,][]{fan06}, which were soon followed by systematic identification of fainter co-eval quasars through surveys using the Canada--France--Hawaii Telescope \citep[e.g.,][]{willott05}.
Subsequently, the sample size at the bright and faint ends has been expanded significantly by the Panoramic Survey Telescope And Rapid Response System 1 \citep[Pan-STARRS1; e.g.,][]{banados14} survey and the Hyper Suprime-Cam \citep[HSC;][]{miyazaki18} Subaru Strategic Program (SSP) survey \citep[e.g.,][]{p1}, 
among others.
Near-IR surveys have pushed the frontier to yet higher redshifts, with the first $z > 7$ quasar being found from the UKIRT Infrared Deep Sky Survey \citep{mortlock11}.
The present redshift record is marked by a quasar at $z = 7.642$ \citep{wang21}, identified from a combined dataset of several optical and near-IR surveys 
including the {\it Wide-field Infrared Survey Explorer} \citep[{\it WISE};][]{wright10} survey.
The advent of new world-leading projects in the coming few years, including the Vera C. Rubin Observatory Legacy Survey of Space and Time \citep[LSST;][]{ivezic19}, {\it Euclid}, and 
the {\it Roman Space Telescope} \citep{spergel15}, is expected to accelerate the exploration further out to $z \sim 10$ \citep[e.g.,][]{euclid19}.

One of the most fundamental quantities characterizing a population of celestial objects is the luminosity function (LF).
In particular, the LF of high-$z$ quasars provides a key piece of information to decode the formation and evolution of supermassive black holes (SMBHs), as well as to measure the quasar contribution to cosmic reionization.
The first attempts to measure the quasar LF at $z \sim 6$ were made with bright SDSS quasars at $M_{1450} < -25$ \citep{fan04, jiang08, jiang09}.
\citet{willott10} included fainter quasars to establish the LF over a broader luminosity range.
This LF has been used as the standard in many subsequent studies, but the constraint on the faint end ($M_{1450} > -24$) was still weak, due to the small number of contributing objects.
A robust measurement of the LF at $z = 6$ down to well below the break luminosity ($-28 < M_{1450} < -21$) was achieved by \citet{p5}, who exploited a large sample of faint quasars drawn from the HSC-SSP survey.
\citet{schindler23} used Pan-STARRS quasars to further improve the bright-end constraint.
At higher redshifts, \citet{venemans13}  measured the cumulative number density using three luminous ($M_{1450} \sim -26$) quasars at $z \ge 6.6$.
\citet{wang19} presented the number densities from 17 quasars at $z \sim 6.7$, measured in three magnitude bins from $M_{1450} \sim -27.5$ to $-25.5$.
However, no other constraints have been reported to date, and thus the overall LF shape has not been determined beyond $z = 6$.

Here we present a measurement of the quasar LF at $z = 7$, based on an unprecedentedly large and complete sample of 35 quasars at $6.55 < z < 7.15$ 
covering a broad luminosity range. 
This is the 19th publication from the Subaru High-$z$ Exploration of Low-Luminosity Quasars \citep[SHELLQs;][]{p1} project, which performs spectroscopic identification and multi-wavelength follow-up observations of high-$z$ quasars drawn from the HSC-SSP imaging data \citep{aihara18}.
Throughout the paper, we use point-spread-function (PSF) magnitudes ($m_{\rm AB}$) and associated errors ($\sigma_{\rm m})$ presented in the AB system \citep{oke83},
corrected for Galactic extinction \citep{schlegel98}.
The cosmological parameters of $H_0$ = 70 km s$^{-1}$ Mpc$^{-1}$, $\Omega_{\rm M}$ = 0.3, and $\Omega_{\rm \Lambda}$ = 0.7 are assumed.

\section{Sample and methods}

\begin{figure}
\epsscale{1.15}
\plotone{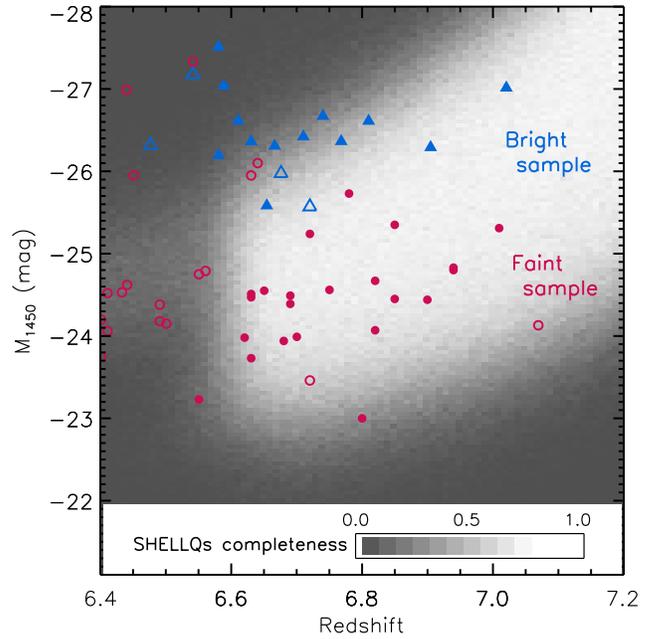}
\caption{Quasars used for the LF determination by \citet[][blue triangles]{wang19} and those discovered by SHELLQs (red circles) over the $z$ -- $M_{1450}$ plane.
The quasars in the present LF sample are represented by the filled symbols, while all the others are represented by the open symbols.
The background colors present completeness of the SHELLQs $z$-dropout selection from 0.0 (dark gray) to 1.0 (white).
\label{fig:completeness}}
\end{figure}

The present measurement combines two quasar samples, called the ``bright sample" and the ``faint sample" hereafter.
They are summarized in Table \ref{tab:sample} and are displayed in Figure \ref{fig:completeness}.
The bright sample is based on 17 quasars at $6.45 < z < 7.05$ presented by \citet{wang19}.
These quasars were identified from an effective area of 13,020 deg$^2$, using the imaging data from Pan-STARRS1, the DESI Legacy imaging Surveys \citep{dey19}, the UKIRT/VISTA Hemisphere Surveys, and the {\it WISE} survey.
We selected 15 quasars at $z > 6.55$ for the present purpose, and further removed two quasars with $y_{\rm AB} > 20.8$ given that the spectroscopic follow-up at fainter magnitudes was incomplete\footnote{
The spectroscopic completeness is 100 \% in all brighter magnitude bins but $20.6 < y_{\rm AB} < 20.7$, where we take the reported completeness of 91 \% into account in the LF calculations.
}
\citep[see Figure 8 of][]{wang19}.


The faint sample comes from SHELLQs, which has so far published the discovery of 162 low-luminosity quasars at $z > 5.6$ \citep{p1,p2,p4,p10,p7,p16}.
The present analysis concerns $z$-band dropout sources with $y$-band detection, the signature of the Ly$\alpha$ break at $z \sim 7$.
Spectroscopic identification has been completed for $z$-dropout candidates selected via a Bayesian probabilistic algorithm in the SSP public data release (DR) 3 fields (an effective area of 957 deg$^2$ with our quality cuts) and for those selected via color cuts in the SSP internal DR S20A fields \citep[779 deg$^2$; see][]{pdr3}.
The details of the quasar selection criteria are described in \citet{p16}.
We use 22 SHELLQs quasars\footnote{
The present analysis does not include type-II quasar candidates, i.e., those objects that show very strong but narrow Ly$\alpha$ emission.
}
at $6.55 < z < 7.05$.
There are a few more SHELLQs quasars in this redshift range, but they do not pass the selection algorithm with the quality flags and imaging properties taken from the latest DR.
For example, the highest-redshift SHELLQs quasar at $z = 7.07$ \citep{p7} and one of the faintest quasars with $z = 6.72$ and $M_{1450} = -23.5$ (see Figure \ref{fig:completeness}) 
happen to fall on regions affected by cosmic ray and a nearby bright star, respectively, 
and are thus excluded from the present measurement.

\begin{deluxetable}{cccc}
\tablecaption{Quasar sample\label{tab:sample}}
\tablehead{
\colhead{Name} & \colhead{Redshift} & \colhead{$M_{1450}$}
} 
\startdata
\multicolumn{3}{c}{Bright sample (Wang et al. 2019)}\\\hline
$J$003836.10$-$152723.6  &     7.02 &    $-27.01$\\
$J$041128.63$-$090749.7   &   6.81  &   $-26.61$\\
$J$070626.38$+$292105.5 &     6.58  &   $-27.51$ \\
$J$082931.98$+$411740.9  &    6.77  &   $-26.36$ \\
$J$083737.83$+$492900.6 &     6.71  &   $-26.42$ \\
$J$083946.88$+$390011.4  &   6.91  &   $-26.29$ \\
$J$091054.54$-$041406.8   &   6.63 &    $-26.36$\\
$J$092347.12$+$040254.6    &  6.61   &  $-26.61$   \\
$J$110421.58$+$213428.9   &   6.74 &    $-26.67$ \\
$J$113508.92$+$501132.6  &    6.58  &   $-26.19$ \\
$J$121627.58$+$451910.7  &    6.65  &   $-25.58$ \\
$J$213233.18$+$121755.2  &    6.59 &    $-27.04$\\
$J$223255.16$+$293032.3 &     6.67 &    $-26.31$\\\hline
\multicolumn{3}{c}{Faint sample (SHELLQs)}\\\hline
  $J$000142.54$+$000057.5    &   6.69  & $-24.49$\\
  $J$011257.84$+$011042.4     &  6.82 & $-24.07$\\
  $J$021316.94$-$062615.2     &  6.72 & $-25.24$\\
  $J$021430.90$+$023240.4    &  6.85 & $-25.35$\\
  $J$021847.04$+$000715.0     &   6.78 & $-25.73$\\
  $J$091041.14$+$005646.3  &   6.65 & $-24.55$\\
  $J$091906.33$+$051235.3    &   6.62 & $-23.98$\\
  $J$102314.45$-$004447.9     &   6.63 & $-24.47$\\
  $J$103537.74$+$032435.7    &   6.63 & $-24.51$\\
  $J$113034.65$+$045013.1    &   6.68 & $-23.94$\\
  $J$120505.09$-$000027.9      &  6.75 & $-24.56$\\
  $J$123137.77$+$005230.3   &    6.69 & $-24.39$\\
  $J$131050.13$-$005054.1     &   6.82 & $-24.67$\\
  $J$131746.92$+$001722.2    &   6.94 & $-24.80$\\
  $J$133833.25$-$001832.9     &   6.70 & $-23.99$\\
  $J$134905.63$+$015608.9     &   6.94 & $-24.83$\\
  $J$140344.28$+$423114.3   &    6.85 & $-24.45$\\
  $J$142903.08$-$010443.4    &   6.8 & $-23.00$\\
   $J$144045.91$+$001912.9   &    6.55 & $-23.23$\\
   $J$145005.39$-$014438.9    &    6.63  & $-23.73$\\
  $J$221027.24$+$030428.5   &    6.9 & $-24.44$\\
  $J$235646.33$+$001747.3    &  7.01 & $-25.31$\\
\enddata
\end{deluxetable}

The completeness, or selection function, of the bright sample has been kindly provided by F. Wang \citep[private communication; see Figure 6 of][]{wang19}. 
For the faint sample, we derived the selection function using exactly the same method
 as described in \citet{p5}, which presents our LF measurement at $z = 6$.\footnote{
Details of the selection conditions have changed over the course of the SHELLQs project, as described in our previous papers. The sample and completeness of the present work are consistently defined with the latest conditions.
}
We repeat only the essence in what follows.
The detection probability of the HSC-SSP imaging for a given magnitude ($ m_{\rm AB}$) has proven to be well approximated by the function
\begin{equation}
f_{\rm det} (m_{\rm AB}) = \frac{1}{2} (\tanh [2.4 (m_{\rm AB}^{5\sigma} - m_{\rm AB})] +1 ),
\end{equation}
where $m_{\rm AB}^{5\sigma}$ is a 5$\sigma$ limiting magnitude measured for every 12\arcmin $\times$ 12\arcmin\ patch of the survey footprint.
SHELLQs selects point sources, whose completeness $f_{\rm ps} (m_{\rm AB})$ has been estimated with HSC-SSP sources observed by 
higher-resolution {\it Hubble Space Telescope} Advanced Camera for Surveys \citep{leauthaud07}; $f_{\rm ps} (m_{\rm AB})$ is always $>$80 \% at the magnitudes we are concerned with ($y_{\rm AB} < 24$ mag).
We created mock high-$z$ quasars based on 319 SDSS spectra of luminous quasars at $z \simeq 3$, by shifting them to higher redshifts and applying Ly$\alpha$ absorption caused by the intergalactic medium \citep[IGM;][]{songaila04, eilers18}.
Each of the mock quasars was assigned random values of redshift, $M_{1450}$, and sky coordinates within the HSC-SSP survey footprint (and thus $m_{\rm AB}^{5\sigma}$), 
providing apparent magnitudes and errors.

We extracted a portion of the mock quasars, such that a quasar with magnitude $y_{\rm AB}$ has probability $f_{\rm det} (y_{\rm AB}) \times f_{\rm ps} (y_{\rm AB})$ of being selected.
The extracted quasars were then fed into the SHELLQs selection algorithm incorporating a Bayesian probabilistic approach and color cuts.
The fraction of selected quasars among all the created mock quasars, as a function of $z$ and $M_{1450}$, provides the completeness.
The reader is referred to \citet{p5} for full details.
The resultant completeness is presented in Figure \ref{fig:completeness}.
It indicates that our $z$-dropout selection is most sensitive to quasars at $z > 6.6$. 
The low completeness at the bright end is due to our requirement of $i$-band non-detection for $z$-dropouts; luminous quasars could have detectable emission even in the \citet{gunn65} trough, and thus would not pass the selection.
Indeed, the two SHELLQs quasars at $z \sim 6.65$ and $M_{1450} \sim -26.0$ (see Figure \ref{fig:completeness}) are not in the present LF sample because of their $i$-band detection.
The completeness remains high at $z > 7.1$, although 
we found no quasars above $z = 7.07$.
While Ly$\alpha$ is well within the HSC $y$-band transmission up to $z \sim 7.5$, it is because (i) the $y$ band is progressively dominated by IGM absorption and (ii) Ly$\alpha$ shifts to the wavelength where the detector sensitivity is low.
The imaging observations could thus detect only luminous quasars, which are too rare to be included in the $\sim$1000 deg$^2$ footprint of the SSP survey.
We set the redshift interval of the present analysis to $6.55 < z < 7.15$, above which the completeness of the bright sample falls close to zero.

\section{Results and Discussion}

\begin{figure*}
\centering
\includegraphics[clip, width=5.5in]{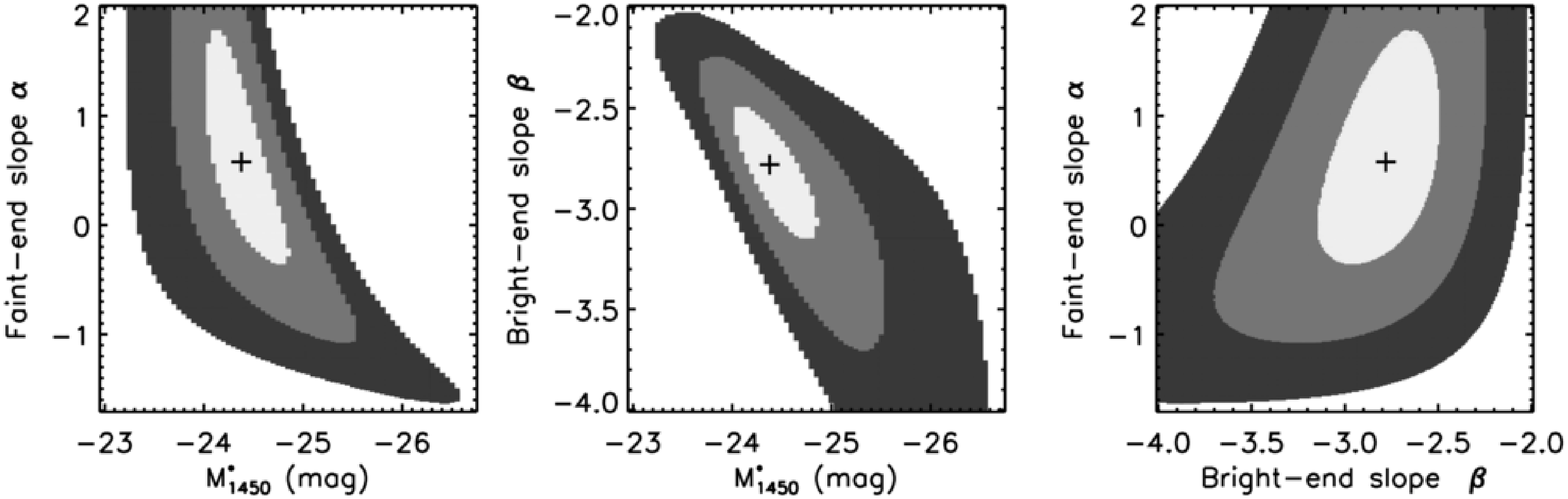}\\
\includegraphics[clip, width=5.5in]{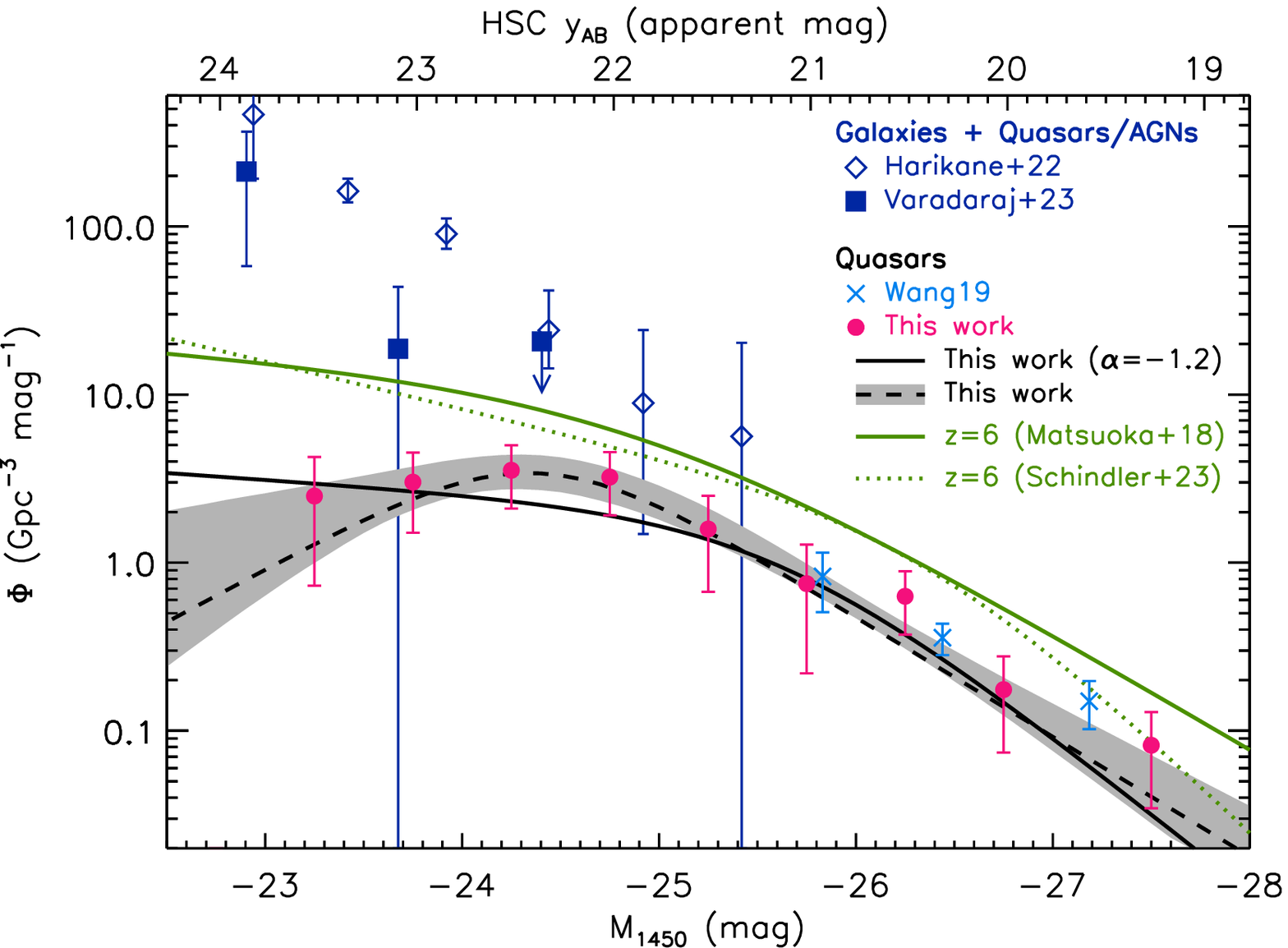}
\caption{Top panels: confidence regions (light gray: 1$\sigma$, gray: 2$\sigma$, dark gray: 3$\sigma$) of the LF parameters.
The best-fit values are marked by the crosses.
Bottom panel: binned LF (red dots) and the standard parametric LF calculated at $z = 6.8$ (solid line; the faint-end slope fixed to $\alpha = -1.2$) from this study.
The dashed line and shaded area represent the best-fit and 1$\sigma$ confidence region, respectively, with $\alpha$ being a free parameter.
For comparison, we also show the binned LF at $z = 6.7$ \citep[][light-blue crosses]{wang19} and the parametric LF measured at $z = 6$ (\citet{p5}, green solid line; \citet{schindler23}, green dotted line).
The blue diamonds and squares represent the binned LF of Lyman break galaxies (AGNs are not excluded) at $z \sim 7$ taken from \citet{harikane22} and \citet{varadaraj23}, respectively
(slight horizontal offsets have been added to improve visibility).
The upper axis gives approximate $y$-band magnitudes for a quasar at $z = 6.8$.}
\label{fig:lf}
\end{figure*}

Figure \ref{fig:lf} and Table \ref{tab:lf} (top part) present the binned LF of the combined bright and faint sample, calculated with the 1/$V_{\rm a}$ method \citep[][see Equations 6 and 7 of \citet{p5}]{avni80} using the selection functions described above.
For comparison, Figure \ref{fig:lf} also shows the LFs of Lyman break galaxies at $z \sim 7$ measured by \citet{harikane22} and \citet{varadaraj23}.
Note that quasars or active galactic nuclei (AGNs) are not excluded from those LFs.
There is some discrepancy between the two galaxy LFs, which \citet{varadaraj23} attribute to contamination of brown dwarfs to the bright end of the \citet{harikane22} sample.
Overall, quasars are a dominant population of sources at $z \sim 7$ for $M_{1450} < -24$ ($y_{\rm AB} < 23$), while galaxies quickly start to outnumber quasars at the fainter magnitudes.

We derived the parametric LF of the quasars assuming a double power-law model:
\begin{equation}
\Phi (M_{1450}) =  \frac{10^{k (z - 7)} \Phi^*} {10^{0.4 (\alpha + 1) (M_{1450} - M_{1450}^*)} + 10^{0.4 (\beta + 1) (M_{1450} - M_{1450}^*)}} ,
\end{equation}
where  $\alpha$ and $\beta$ are the faint- and bright-end slope, respectively, and $M^*_{1450}$ is the break magnitude.
We determined the three parameters along with the normalization factor $\Phi^*$ with a maximum likelihood fit \citep[][see Equation 9 of \citet{p5}]{marshall83}.
We adopt the redshift-evolution slope of $k = -0.78$, which was measured between $z \sim 6$ and $z \sim 6.7$ by \citet{wang19}. 
The best-fit parameters are reported in Table \ref{tab:lf} (the first row of the bottom part), which yields the parametric LF presented in Figure \ref{fig:lf} (dashed line).
An alternative case with $k = -0.70$ \citep{jiang16} was also tested but the results remain almost unchanged.
It is remarkable that the parametric LF shows a positive faint-end slope ($\alpha > -1$; the number density declines toward the faint end), although a negative slope ($\alpha < -1$) is also allowed within the 2$\sigma$ confidence interval.

It is unlikely that our selection misses the faint-end quasars because of the extended emission from the host galaxies. 
Our criterion of point-source selection for $z$-dropouts is very conservative, $0.6 < \mu$/$\mu_{\rm PSF} < 3.0$, where $\mu$ and $\mu_{\rm PSF}$ represent the $y$-band image adaptive moment of a given source and that of the PSF model, respectively.
That is, we allow the sources that are up to three times larger than the PSF size to remain in the candidates for spectroscopic confirmation.
In reality, none of the SHELLQs quasars in the present sample has $\mu$/$\mu_{\rm PSF} > 1.7$.
All but a few of the $\sim$200 Lyman break galaxies at $z \sim 7$ with similar luminosities ($M_{1450} < -22$), drawn from the same HSC-SSP images \citep{harikane22}, also meet $\mu$/$\mu_{\rm PSF} < 3.0$, even without the central point source (i.e., AGN).

A positive slope, if confirmed, might indicate that the obscured fraction increases and/or the mass-accretion efficiency decreases toward the low-mass end of the SMBH mass distribution at $z \ge 7$.
Indeed, observations at lower redshifts ($z < 4$) suggest that the obscured fraction increases significantly toward lower (intrinsic) luminosities \citep[e.g.,][]{merloni14, ueda14, toba21}.
There is also an indication of a higher obscured fraction at higher redshifts up to $z \sim 6$ \citep[e.g.,][]{vito18, gilli22}.
We have identified 23 candidate obscured quasars at $z \sim 6$ from SHELLQs \citep[e.g.,][]{p16}, based on the extremely luminous and narrow Ly$\alpha$ emission.
\citet{onoue21} detected strong C IV $\lambda$1549 emission in one of them, pointing to the presence of hard ionizing radiation from AGN.
A number of other observations \citep[e.g.,][]{fujimoto22, endsley23} are revealing (candidates for) such a high-$z$ obscured population, and substantially more objects may be identified in the coming few years with the {\it James Webb Space Telescope} ({\it JWST}).

On the other hand, at lower redshifts ($z \le 6$), the faint-end slope is close to flat but is always negative.
For example, the measurements based on the HSC-SSP data suggest $\alpha = -1.30 \pm 0.05$ at $z = 4$ \citep{akiyama18}, $\alpha = -1.22^{+0.03}_{-0.10}$ at $z = 5$ \citep{niida20}, and $\alpha = -1.23^{+0.44}_{-0.34}$ at $z = 6$ \citep{p5}.
We thus performed another set of parametric fittings with a fixed slope of $\alpha = -1.2$. 
This case actually lies within the $\sim$2$\sigma$ confidence region when all four LF parameters are varied above, and indeed, the best-fit line with $\alpha = -1.2$ gives a reasonable agreement with the binned LF (see Figure \ref{fig:lf}).
Therefore, we regard the case with $\alpha = -1.2$ (the third row in the bottom part of Table \ref{tab:lf}) as our standard fit to the quasar LF at $z = 7$,
and use the corresponding parameters in the following discussions.
For reference, the table also lists the fitting results when two parameters, $\alpha$ and $M_{1450}$, are fixed to the values measured at $z = 6$ \citep{p5}.

\begin{deluxetable*}{cccccl}
\tablecaption{Quasar luminosity function at $z = 7$\label{tab:lf}}
\tabletypesize{\scriptsize}
\tablehead{
}
\startdata
\multicolumn{6}{c}{Binned luminosity function}\\\hline
$M_{1450}$ & $\Delta M_{1450}$ & $\Phi  (M_{1450})$ & $N_{\rm obj}$ & $V_{\rm a}$ \\\hline
 $-23.25$ & 0.5 &        2.5  $\pm$ 1.8    &  2 & 1.607\\
 $-23.75$ & 0.5 &        3.0  $\pm$ 1.5   &  4 & 2.656 \\
 $-24.25$ & 0.5 &       3.5  $\pm$ 1.4       &  6 & 3.386 \\
 $-24.75$ & 0.5 &       3.2  $\pm$ 1.3       &  6 & 3.719 \\
 $-25.25$ & 0.5 &       1.58  $\pm$ 0.91    &  3 & 3.786 \\
 $-25.75$ & 0.5 &       0.75 $\pm$  0.53    &  2 & 5.336 \\
 $-26.25$ & 0.5 &      0.63 $\pm$  0.26     &  6 & 19.034 \\
 $-26.75$ & 0.5 &       0.18 $\pm$  0.10    &  3 & 34.171 \\
 $-27.50$ & 1.0 &       0.082 $\pm$  0.047   &  3 & 36.563\\\hline\hline
\multicolumn{6}{c}{Parametric luminosity function}\\\hline
$\Phi^*$ & $M_{1450}^*$ & $\alpha$ & $\beta$ & $k$ & comment\\\hline
$4.75^{+1.14}_{-1.36}$ & $-24.38^{+0.33}_{-0.47}$ & $0.58^{+1.20}_{-0.93}$ & $-2.78^{+0.28}_{-0.36}$ & $-0.78$ (fixed) & free $\alpha$\\
$4.90^{+1.27}_{-1.43}$ & $-24.40^{+0.35}_{-0.45}$ & $0.52^{+1.24}_{-0.89}$ & $-2.79^{+0.30}_{-0.35}$ & $-0.70$ (fixed) & free $\alpha$, different $k$\\
$1.35^{+0.47}_{-0.30}$ & $-25.60^{+0.40}_{-0.30}$ & $-1.20$ (fixed)               & $-3.34^{+0.49}_{-0.57}$ & $-0.78$ (fixed) & standard\\
$2.07^{+0.17}_{-0.20}$ & $-24.90$ (fixed)                & $-1.23$ (fixed)               & $-2.72^{+0.21}_{-0.22}$ & $-0.78$ (fixed) & fixed $M_{1450}^*$\\
\enddata
\tablecomments{$M_{1450}$ and $\Delta M_{1450}$ denote the center and width of a magnitude bin, respectively, while
	$N_{\rm obj}$ represents the number of quasars contained in the bin.
	The number densities $\Phi  (M_{1450})$ and $\Phi^*$ (normalization at $z = 7$) are given in units of Gpc$^{-3}$ mag$^{-1}$.
	$V_{\rm a}$ (Gpc$^3$) represents the cosmic volume available to discover quasars in the present sample.}
\end{deluxetable*}

\begin{figure}
\epsscale{1.15}
\plotone{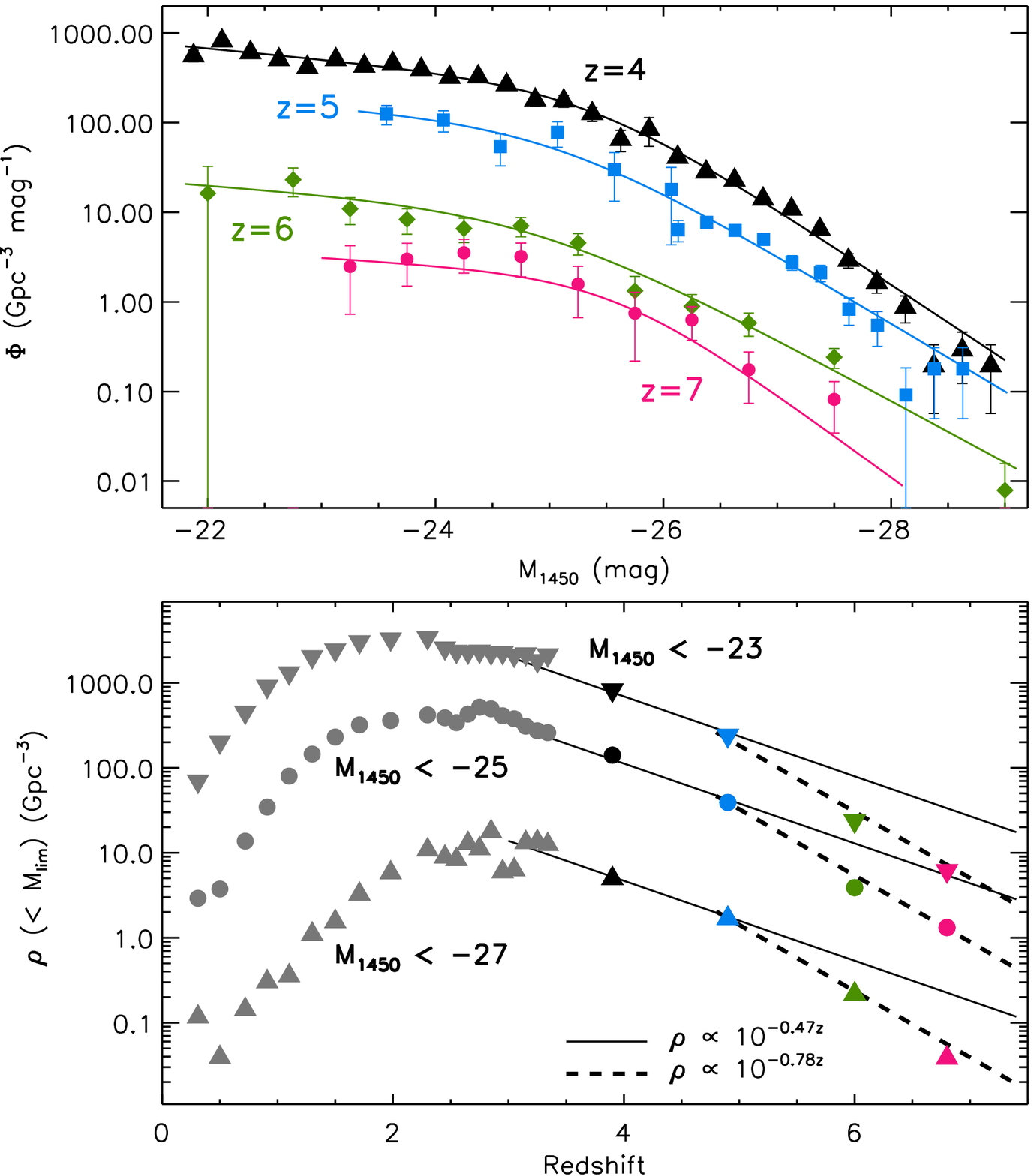}
\caption{Evolution of the quasar LF.
Top panel: binned and parametric LF of quasars at $z = 4$ \citep[black triangles and line;][]{akiyama18}, 
$z = 5$ \citep[blue squares and line;][]{niida20},
$z = 6$ \citep[green diamonds and line;][]{p5},
and $z = 7$ (red circles and line; this work).
Bottom panel: integrated number density of quasars with $M_{1450} < -23$ (inverse triangles), $M_{1450} < -25$ (dots), and $M_{1450} < -27$ (triangles).
The data at $z < 3.5$ are taken from \citet{kulkarni19}.
Density evolution laws ($\rho \propto 10^{-k z}$) with $k = -0.47$ and $k = -0.78$ are represented by the solid and dashed lines, respectively, with arbitrary scaling.
\label{fig:LFevol}}
\end{figure}

Figure \ref{fig:LFevol} (top panel) compares the quasar LFs at redshifts $z = 4$, $5$, $6$, and $7$.
The overall shape remains remarkably similar through this redshift range,\footnote{
The parametric LFs at z = 6 and z = 7 have the same faint-end slope by design, but the overall similarity in shape is also observed in the binned LFs at the two redshifts.}
while the amplitude decreases significantly toward higher redshift.
Figure \ref{fig:LFevol} (bottom panel) displays this decreasing trend more explicitly, 
with the data at $z < 3.5$ supplemented from \citet{kulkarni19}.
The density peaks at lower redshifts for lower-luminosity quasars, an effect which is termed ``down-sizing of AGNs" \citep{ueda03, barger05}.
The cumulative number density declines rapidly toward higher redshifts following $\rho \propto 10^{-k z}$ with $k = -0.47$ \citep{fan01} at $3 < z < 5$, and then it drops even 
faster at $z > 5$.
\citet{jiang16} reported $k = -0.72 \pm 0.11$ at $5 < z < 6$, and \citet{wang19} found a steeper slope $k = -0.78 \pm 0.18$ at $6 < z < 7$.
While the above measurements were limited to the most luminous quasars ($M_{1450} < -26$ mag), Figure \ref{fig:LFevol} suggests that the density of
low-luminosity quasars ($M_{1450} < -23$) is also consistent with $k = -0.78$ at $5 < z < 7$.
The similar epoch of turnover (i.e.,  to accelerating decline toward the highest redshifts) at different luminosities may indicate that the density evolution is governed by different physics from that drives the down-sizing at lower redshifts.
Comparison of the present LF measurements with theoretical models \citep[e.g.,][]{li22, oogi22} will be a subject of future papers.

Finally, we estimate the quasar contribution to cosmic reionization at $z = 7$, using the LF estimated above.
Section 5 of \citet{p5} describes the details of this calculation.
The ionizing photon density, $\dot{n}_{\rm ion}$, is calculated assuming that quasar spectra follow a broken power law \citep{lusso15} 
and that a photon escape fraction is unity.
By integrating our standard LF over $-30 < M_{1450} < -18$, we get $\dot{n}_{\rm ion} = 10^{48.2 \pm 0.1}$ s$^{-1}$ Mpc$^{-3}$. 
This result is insensitive to the integrated magnitude range, since the LF is close to flat at the faint end and predicts a very small number of objects at the bright end.
The value of the ionizing photon density that would balance the rate of recombination at $z = 7$ is given by $\dot{n}_{\rm ion} = 10^{50.2} C_{\rm H II}$ (s$^{-1}$ Mpc$^{-3}$), where $C_{\rm H II}$ represents an effective \ion{H}{2} clumping factor \citep{madau99, bolton07}.
The observed $\dot{n}_{\rm ion}$ is thus less than 1 \% of the critical density for any plausible value of $C_{\rm H II} \ge 1$, clearly suggesting that quasars cannot be a major contributor to the reionization of the universe at this redshift.

The biggest original goal of the SHELLQs project was to establish the quasar LF at $6 \le z \le 7$, which is now completed.
The faint sample used here was drawn from the HSC-SSP data contained in the public DR 3, which have been acquired over 278 out of the 330 nights allocated to the SSP.
The final DR, expected to happen in Fall 2023, will contain the data taken over the remaining 52 nights;
we may use those data to further improve LF constraints, though it is not expected to increase the sample size significantly.
In the meantime, the unprecedented sensitivity of {\it JWST} is starting to find signatures of low-mass AGNs in high-$z$ galaxies \citep[e.g.,][]{harikane23, larson23, onoue23},
and is also expected to shed new light on obscured side of the early AGN activity \citep[e.g.,][]{kocevski23}. 
With the advent of new cutting-edge facilities including the Rubin LSST, {\it Euclid}, and {\it Roman}, observations in the coming ten years will 
bring crucial constraints 
on the quasar/AGN LF out to very high redshifts, and thus will provide a key to discern models of the birth and growth of SMBHs throughout the epoch of reionization.

\acknowledgments

This research is based on data collected at Subaru Telescope, which is operated by the National Astronomical Observatory of Japan. 
We are honored and grateful for the opportunity of observing the Universe from Maunakea, which has the cultural, historical and natural significance in Hawaii.
We appreciate the staff members of the telescope for their support during our FOCAS observations.

This work is also based on observations made with the GTC, installed at the Spanish Observatorio del Roque de los Muchachos 
of the Instituto de Astrof\'{i}sica de Canarias, on the island of La Palma.
We thank Stefan Geier and other support astronomers for their help during preparation and execution of our observing program.

Y. M. was supported by the Japan Society for the Promotion of Science (JSPS) KAKENHI Grant No. JP17H04830, No. 21H04494, and the Mitsubishi Foundation grant No. 30140.
M. O and K. I are supported by the National Natural Science Foundation of China (12073003, 11991052, 11721303, 12150410307, 11950410493)
and the China Manned Space Project Nos. CMS-CSST-2021-A04 and CMS-CSST-2021-A06.
K. I. acknowledges support by grant PID2019-105510GB-C33 funded by MCIN/AEI/10.13039/501100011033 and ``Unit of excellence Mar\'ia de Maeztu 2020-2023" awarded to ICCUB (CEX2019-000918-M).

The HSC collaboration includes the astronomical communities of Japan and Taiwan, and Princeton University.  The HSC instrumentation and software were developed by the National Astronomical Observatory of Japan (NAOJ), the Kavli Institute for the Physics and Mathematics of the Universe (Kavli IPMU), the University of Tokyo, the High Energy Accelerator Research Organization (KEK), the Academia Sinica Institute for Astronomy and Astrophysics in Taiwan (ASIAA), and Princeton University.  Funding was contributed by the FIRST program from the Japanese Cabinet Office, the Ministry of Education, Culture, Sports, Science and Technology (MEXT), the Japan Society for the Promotion of Science (JSPS), Japan Science and Technology Agency  (JST), the Toray Science  Foundation, NAOJ, Kavli IPMU, KEK, ASIAA, and Princeton University.
 
This paper is based on data collected at the Subaru Telescope and retrieved from the HSC data archive system, which is operated by Subaru Telescope and Astronomy Data Center (ADC) at NAOJ. Data analysis was in part carried out with the cooperation of Center for Computational Astrophysics (CfCA) at NAOJ.  
 
This paper makes use of software developed for Vera C. Rubin Observatory. We thank the Rubin Observatory for making their code available as free software at http://pipelines.lsst.io/. 
 
The Pan-STARRS1 Surveys (PS1) and the PS1 public science archive have been made possible through contributions by the Institute for Astronomy, the University of Hawaii, the Pan-STARRS Project Office, the Max Planck Society and its participating institutes, the Max Planck Institute for Astronomy, Heidelberg, and the Max Planck Institute for Extraterrestrial Physics, Garching, The Johns Hopkins University, Durham University, the University of Edinburgh, the Queen's University Belfast, the Harvard-Smithsonian Center for Astrophysics, the Las Cumbres Observatory Global Telescope Network Incorporated, the National Central University of Taiwan, the Space Telescope Science Institute, the National Aeronautics and Space Administration under grant No. NNX08AR22G issued through the Planetary Science Division of the NASA Science Mission Directorate, the National Science Foundation grant No. AST-1238877, the University of Maryland, Eotvos Lorand University (ELTE), the Los Alamos National Laboratory, and the Gordon and Betty Moore Foundation.



\end{document}